# Experimental determination of conduction channels in atomic scale conductors based on shot noise measurements

Ran Vardimon, Marina Klionsky, and Oren Tal

Department of Chemical Physics, Weizmann Institute of Science, Rehovot 76100, Israel

We present an experimental procedure for obtaining the conduction channels of low-dimensional conductors based on shot noise measurements. The transmission coefficient for each channel is determined numerically from the measured conductance and Fano factor. The channel analysis is demonstrated for atomic contacts of Ag, Au, Al and Pt, showing their channel evolution as a function of conductance and mechanical elongation. This approach can be readily applied to map the conduction channels in a wide range of nanoscale conductors under different conditions.

The conductance of a coherent quantum conductor can be described as a sum of independent conduction channels, originating from the quantization of the electron wave modes within the conductor[1]. The conductance associated with each channel is limited by the conductance quantum $G_0 = 2e^2/h$ while the total conductance is given by $G = G_0 \sum_{n=1}^{N} \tau_n$ where $0 \leq \tau_n \leq 1$ are the transmission coefficients of N conduction channels. This description is related to a wide variety of quantum systems such as mesoscopic scale quantum dots[2], nanowires and nanotubes[3,4] as well as atomic contacts[5] and molecular junctions[6]. The set of coefficients $\{\tau_n\}_1^N$ can be viewed as a unique "PIN code" which describes the transport properties of the conductor. The possibility to measure the PIN code rather than merely the overall conductance would enable one to link between the observed conductance to the fundamental principles of electronic transport. For example, by analyzing the sub-gap structure of current-voltage curves in the superconducting state, Scheer and co-workers demonstrated that the number of channels in single atom contacts is limited by the number of atomic valence orbitals[7,8]. Experimental studies of conduction channels have shed light on the orbital structure[9,10], atomic configurations[11] and electron-vibration interactions[12] in atomic scale junctions. Still, up to date experimental data regarding the conduction channels is limited since obtaining the transmission coefficients has proven to be very challenging. While the transmission coefficients can be calculated from the sub-gap structure, these measurements are restricted to systems that can be driven to the superconducting state. For non-superconducting systems, several experiments on single-molecule junctions showed that the transmission coefficients can be estimated from shot noise measurements[12–14]. Although shot noise can be measured in a wide variety of systems, analytical treatment of shot noise data allows calculating the transmission coefficients only when the conductance is composed of at most two channels.

In this paper, we describe a straightforward approach to determine the conduction channels from shot noise measurements beyond the analytical limit of two channels. Our approach is based on enumerating the set of transmission coefficients and identifying the range of results which are consistent with the measured shot noise and conductance. Using a break junction setup to form Ag, Au, Al and Pt atomic contacts we demonstrate how different aspects of electronic transport are revealed by numerically obtaining the channel resolution. A statistical analysis of transmission coefficients obtained for ensembles of Ag and Au contacts reveals a consistent tendency of the conduction channels to fully saturate one by one as the conductance of the contact is increased. The channel distribution obtained for Al contacts proves to be in excellent agreement with both sub-gap measurements and theoretical calculations. Finally, we show that the number of conduction channels through single atom Pt contacts can vary significantly between two different contact configurations.

Combining shot noise and conductance data can provide information on the conduction channels. The conductance is driven from the average current, often referred to as the first moment of current. Distinctively, shot noise describes the second moment of current and therefore provides independent information[15]. The suppression ratio of shot noise from Poissonian noise is expressed by the Fano factor $F = \sum_{n=1}^{N} \tau_n(1-\tau_n) / \sum_{n=1}^{N} \tau_n$, which can be determined experimentally from the dependence of shot noise on applied bias voltage (see supplementary material[16]). Therefore, the measured Fano factor (F) and conductance (G) provide two independent equations for the transmission coefficients. These expressions can be used to determine the coefficients $\{\tau_n\}_1^N$ when there are up to N=2 conduction channels, however, for more than two channels, the exact set of transmissions cannot be uniquely determined.

In general, a given combination of F and G can result from an infinite number of possible sets of transmission coefficients. However, certain values of F and G may limit the range of possible $\tau_n$ values. Without loss of generality, we assume the coefficients are ordered by decreasing transmission $\tau_1 \geq \tau_2 \geq ... \geq \tau_N$. Using a finite precision $\Delta \tau$ for the transmission coefficients limits the number of possible transmission sets to a finite number. Our procedure enumerates the $(1/\Delta \tau)^N$ possible sets. For each set, the Fano factor and conductance are computed and compared to the experimental values of F and G. A transmission set $\{\tau_n\}_1^N$ is considered to match the experimental values if it satisfies the following inequalities:

$$(1)\ G - \Delta G \leq G_0 \sum_{n=1}^{N} \tau_n \leq G + \Delta G$$

$$(2)\ F - \Delta F \leq \frac{\sum_{n=1}^{N} \tau_n(1-\tau_n)}{\sum_{n=1}^{N} \tau_n} \leq F + \Delta F$$

where ΔF and ΔG are the experimental errors in F and G, respectively. We define $\{\tau_{i,n}\}_{n=1}^{N}$ as the i'th set out of k transmission sets that match the experimental values. The transmission coefficient $\tau_n$ can now be determined to be in the range between $\tau_n^{min} = \min\{\tau_{i,n}\}_{i=1}^{k} - \Delta\tau$ and $\tau_n^{max} = \max\{\tau_{i,n}\}_{i=1}^{k} + \Delta\tau$. The additional margins $\Delta\tau$ are added to ensure all possible solutions for $\tau_n$ are included between $\tau_n^{min}$ and $\tau_n^{max}$ (more details can be found in the supplementary material [16]). Note that spin-degeneracy is assumed in this analysis, however it is straightforward to extend the method to account for non-degenerate systems.

The extraction of the transmission coefficients is illustrated in Fig. 1 for a certain combination of F=0.20±0.01 and G=2.00±0.01. The panel for each coefficient $\tau_n$ (n=1-4) displays the distribution $\{\tau_{i,n}\}_{i=1}^{k}$ of the computed values that match the given F and G. While in principle there are an infinite number of solutions, the histograms of the individual coefficients that match the given F and G imply that $\tau_n$ do not span the whole range between 0-1. Rather, they are limited to a much smaller range. As a consequence, each transmission coefficient can be determined up to a limited uncertainty, which is given by the edges of the corresponding distribution (Fig. 1, rightmost panel). The resulting uncertainties in $\tau_n$ greatly depend on the measured F and G values. Detailed information on the sensitivity with respect to F and G can be found in the supplementary material[16]. As will be demonstrated below, in many cases the transmission coefficients of nanoscale conductors can be determined with sufficient accuracy to capture the important physical aspects of the conduction channels.

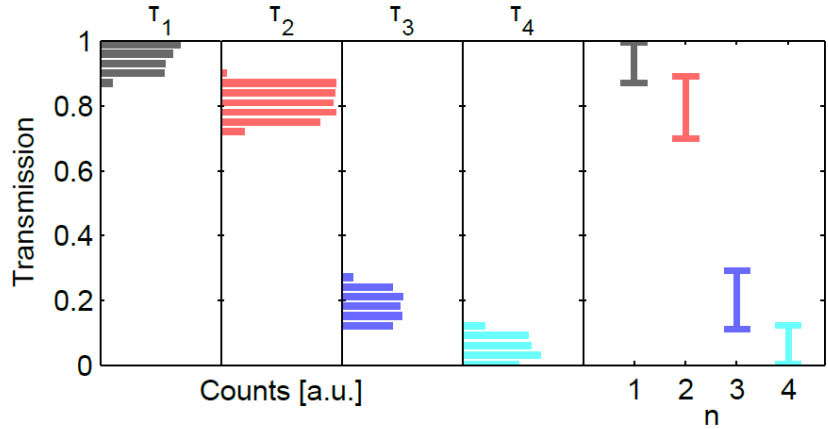

FIG. 1. Calculated transmission coefficients for four channels matching F=0.20±0.01 and G=2.00±0.01, using N=4 and Δτ=0.01. The four left panels show the distribution of calculated values for τ₁-τ₄. The rightmost panel shows the transmission range for each coefficient: τ ₁=0.86-1.00, τ ₂=0.69-0.90, τ ₃=0.10-0.30, τ ₄=0.00-013.

The presented analysis was used to examine the evolution of conduction channels in atomic contacts under variation of the contact geometry. Atomic contacts were formed at 4.2K in cryogenic vacuum conditions using a mechanical controllable break junction (MCBJ) technique[17]. A microscopic wire (0.1mm diameter) with a small notch at its center is pulled apart in a controlled fashion. During the elongation process the number of atoms in the cross-section of the wire constriction is gradually reduced until reaching a single atom contact. A three point bending mechanism driven by a piezoelectric element was used to control the elongation of the contact with sub-Angstrom resolution. The wires that were used are composed of pure (> 99.99%) Ag, Au, Al and Pt. The differential conductance (dI/dV) of the contacts was measured across the wire using a lock-in technique. In order to measure shot noise, two sets of voltage amplifiers were connected in parallel to the sample, and the cross-spectrum of the two signals was calculated as function of applied bias current[18]. Detailed information regarding shot noise measurements and the extraction of the Fano factor can be found in the supplementary material[16]. We then apply our numerical method to extract the conduction channels from the measured values F±ΔF, G±ΔG. The transmission coefficients are calculated using $\Delta\tau = 0.005$ and a maximum of $N = 6$ channels, which is a valid high limit since the number of channels for a single atom contact is limited by the number of atomic valence orbitals[7]. In general, the analysis can be repeated for N+1 channels to verify the validity of the results for N channels.

We start by examining the monovalent metals Ag and Au. For these metals, the conductance of a single atom is expected to be carried by one conduction channel, explained by the dominant contribution of a single s valence orbital[7]. For each metal, shot noise measurements were performed on more than 600 distinct atomic contacts. Before each measurement, the metal wire is reformed up to a conductance of $70G_0$ and pulled apart to reach a new atomic configuration. Figures 2(a) and 2(b) show the distribution of F and G values measured on Ag and Au contacts, respectively. The results show a clear tendency to follow the minimum Fano factor curve (represented in both figures as a black line). At integer conductance values, the majority of measurements show a strong suppression of the Fano factor. This suppression is a clear signature for quantized conductance through fully transmitting channels[8,18]. For both metals a full suppression can be observed at $1G_0$ and $2G_0$. However, while at $3G_0$ the Fano factor is strongly suppressed in the case of Ag, a deviation to slightly higher values is observed for Au.

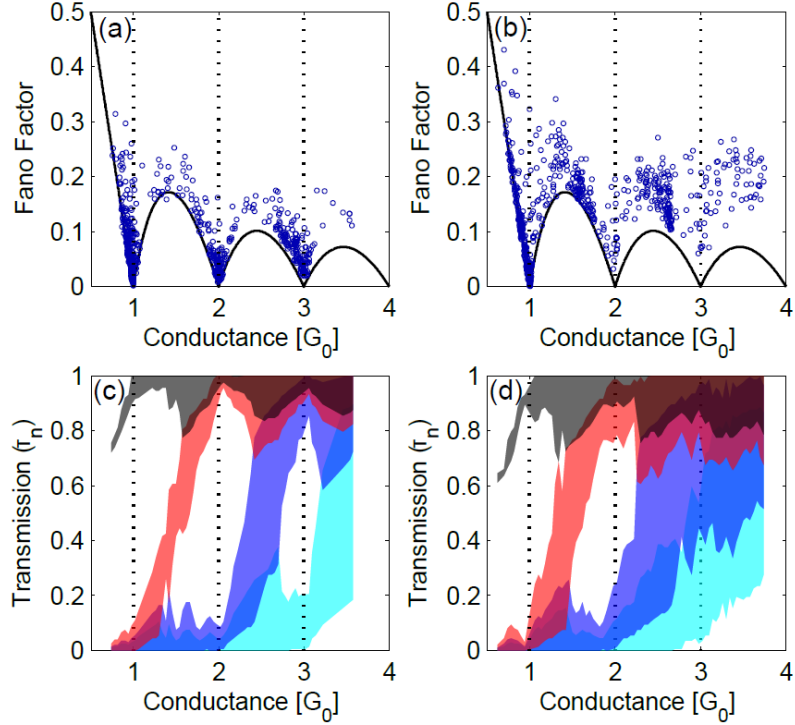

FIG. 2. (a,b) F and G values measured on an ensemble of 633 Ag (a) and 702 Au (b) atomic contacts. The experimental errors do not exceed $\Delta G<0.015 G_0$ and $\Delta F<0.02$. Black line reflects the theoretical minimum Fano factor. (c,d) Computed transmission ranges for $\tau_1$-$\tau_4$ extracted from the data in (a) and (b), respectively. Black, red, blue and cyan areas represent the transmission ranges for $\tau_1$, $\tau_2$, $\tau_3$ and $\tau_4$, respectively.

The numerical analysis of the sets of data that appear in Fig. 2(a) and 2(b) provides the range of values for the transmission coefficients. The coefficients are calculated to account for the statistical spread of the data in the following way. First, the Fano factor values are binned by conductance intervals of $0.04 G_0$. Then, for each bin, the values of $F$ and $\Delta F$ used in eq. (1) and (2) are set to $F=(F_{min}+F_{max})/2$ and $\Delta F=(F_{max}-F_{min})/2$, where $F_{min}$ and $F_{max}$ are the minimum and maximum Fano factor measured within the conductance bin. This ensures that the computed $\tau_n^{min}$ and $\tau_n^{max}$ capture all the possible transmission coefficients that correspond to the Fano factor values within the relevant conductance bin.

Figures 2(c) and 2(d) display the distribution of $\tau_n$ as function of conductance for Ag and Au, respectively. The colored areas show the possible values for $\tau_n$ accounting for both the statistical spread in the Fano factor and the uncertainty in the numerical calculation. The transmission coefficients calculated for Ag (Fig. 2(c)) show that the conduction channels fully open, one by one, as the conductance of the contact increases. Around half integer conductance values (i.e. $1.5 G_0$, $2.5 G_0$, and $3.5 G_0$) the spread in the transmission range increases as a result of the limited accuracy of our analysis at these conductance values (see supplementary material [16]). The spread is lower near integer conductance values, indicating that the conductance is composed of fully open channels. In the case of Au, a clear tendency for sequential channel saturation is also observed up to ~2.5$G_0$, however at higher conductance values this behavior is less pronounced. Altogether, this analysis clearly demonstrates for both noble metals that the conductance is composed of saturated conduction channels and, in the case of non-integer conductance, an additional partly transmitting channel.

The possibility for sequential saturation of conduction channels was suggested before to explain the observed shot noise[8] and conductance fluctuations[19] in monovalent metals, although the transmission coefficients were not obtained directly. Conversely, sub-gap measurements on Au contacts showed that when G>1$G_0$ the conductance can be composed of several partly-opened channels[7,20]. The later experiments were carried out by inducing superconductivity on thin Au layers placed on Al contacts using the proximity effect. The discrepancy from sub-gap measurements could be explained by the difficulties in fitting the IV characteristics in the non-ideal case of superconductivity induced by the proximity effect[21]. Note that for bare Al contacts our results are in good agreement with sub-gap measurements, as will be shown.

A theoretical treatment of the transmission coefficients for Ag and Au was performed by a combination of molecular dynamics simulations and tight-binding calculations[20,22,23]. For both metals, a single channel contributing up to 1$G_0$ is found in agreement with our results, however the sequential saturation behavior is not reproduced and several partially opened channels can be found above 1$G_0$. Interestingly, our results are in very good agreement with a description of free electrons with disorder, in which the atomic neck is modeled by a smooth constriction[24,25]. This is a surprising observation since one would expect that the transmission coefficients will depend on the details of the atomic configuration, rather than only on the diameter of the smallest cross-section.

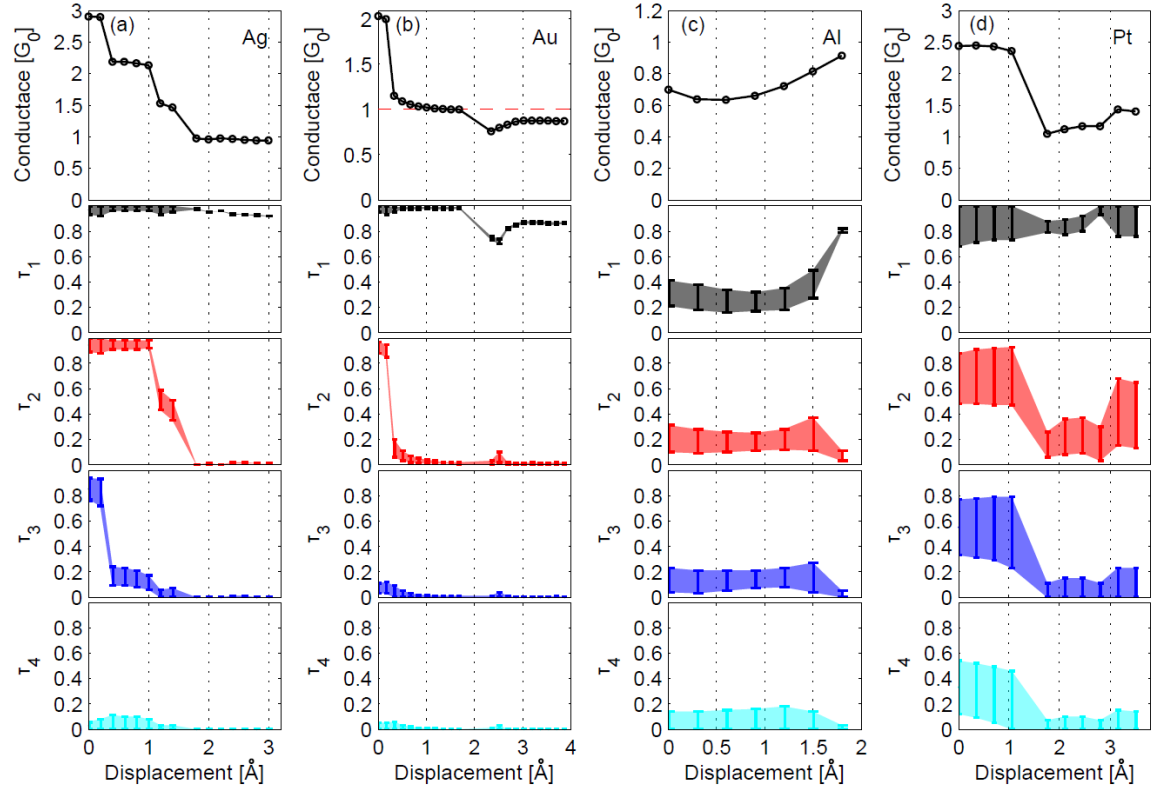

FIG. 3. Conductance traces (black circles) and transmission coefficients of the different channels measured on Ag (a), Au (b), Al (c) and Pt (d) point contacts. Black, red, blue and cyan error bars represent the possible transmission ranges for $\tau_1$, $\tau_2$, $\tau_3$ and $\tau_4$, respectively.

In order to study the effect of the atomic configuration, the transmission coefficients were obtained simultaneously during gradual narrowing of atomic contacts subject to mechanical stretching. Figure 3(a) presents an example trace for Ag, showing the evolution of the conductance (top panel) and transmission coefficients (lower panels) as a function of contact elongation. During elongation, the conductance exhibits a series of plateaus, which are related to stable atomic configurations, separated by sudden jumps due to plastic deformation[26]. The decomposition to transmission coefficients shows that the conductance of each plateau consists of saturated channels accompanying up to one partly open channel. The jumps do not necessarily result in a reduction of the number of channels, as can be seen in the first jump between $3G_0$ to $2.2G_0$, which results from a decrease in the transmission of $\tau_3$. In the presented Au trace (Fig. 3(b)), the last plateau begins with conductance slightly higher than $1G_0$ that gradually decreases to $1G_0$ at 1.8Å. Resolving the conductance in terms of channels provides insight into this observation – throughout the elongation the main channel ($\tau_1$) is fully open while a second channel ($\tau_2$) shows a tunneling-like exponential dependence on electrode displacement. This indicates that tunneling between neighboring atoms can contribute to the overall conductance.

We now turn to study the sp metal Al, where the conductance of a single atom contact is composed by three main channels[9,10]. An example for an Al trace showing the evolution of the conduction channels under elongation of a single atom contact is presented in Fig. 3(c). At the beginning of the trace, the conductance is mainly composed of three approximately equally distributed channels. As the contact is stretched, one channel increases significantly while the other two decrease. This is a typical behavior observed in 15 Al traces examined in this work. The number of observed channels and their dependence on elongation are in very good agreement with previous experimental results obtained using sub-gap measurements[9]. In contrast to Au contacts, in which the superconducting state was induced by the proximity effect, for pure Al contacts the comparison between the two experimental methods is more straightforward[27].

Finally, we examine the conduction channels in atomic Pt contacts. Figure 3(d) shows an example for a conductance trace measured during the elongation of a single atom constriction. After an elongation of ~1A, the conductance shows a sudden drop from $2.5G_0$ to a semi-plateau ranging between $1-1.5G_0$. Similar conductance drops were associated with a geometric reordering of the contact from a single apex atom to a dimer configuration[22,28]. After the transition, the number of observed channels drops from 3-4 channels down to 2 channels, with a possible small contribution from other channels. A similar behavior is observed in more than 30 measured traces. In all traces a highly transmitting channel ($0.7 \leq \tau_1 \leq 1$) is observed during the whole elongation process. Interestingly, the main channel is less affected during the conductance drop while the other channels decrease significantly. These observations are in accordance with theoretical calculations[22,29], which predict a dominant conduction channel associated with a hybridization of s, $p_z$ and $d_{z^2}$ orbitals, and two less transmitting channels, dominated by contributions from $d_{zx}$ and $d_{yz}$ orbitals. The significant spherical symmetry of the main channel could be the reason for its relative insensitivity to the structural changes occurring with elongation.

To conclude, we have presented a numerical method for obtaining the conduction channels of low dimensional conductors from shot noise measurements. Using this analysis, we overcome the analytical limitation for multi-channel determination by relaxing the requirement for exact identification of the transmission coefficients. We have demonstrated the analysis of conduction channels based on shot noise measurements for several metallic contacts. The calculated distribution of channels is in good agreement with sub-gap measurements for Al contacts and for single atom Au contacts, although an unexpected discrepancy is found for larger Au contacts. The obtained channel distribution during the evolution of Ag, Au, Al and Pt contacts exemplify that the conduction channel resolution can shed light on the electronic transport behavior in various systems. We stress that the presented procedure is universal and can, in principal, be applied to any low dimensional conductor.

The authors are grateful to Jan M. van Ruitenbeek and Manohar Kumar (Leiden University) for their valuable support in constructing the experimental setup. The authors kindly thank Sagi Hed and Ophir Samson for their assistance with the calculations and critical readings of the manuscript. O.T thanks the Harold Perlman Family for their support and acknowledges funding by the Israel Science Foundation Grant No. 1313/10, the German-Israeli Foundation Grant No. I-2237-2048.14/2009 and the Minerva Foundation Grant No. 711136.

# Experimental determination of conduction channels in atomic scale conductors based on shot noise measurements

## Supplementary Information

Ran Vardimon, Marina Klionsky, and Oren Tal

**Table of contents**

## S1. Experimental details

The electronic measurements are performed using two switchable circuits: one for conductance measurements (Fig. S1, blue) and a second for measuring electronic noise (red). The conductance circuit consists of a 24 bit NI PCI-4461 DAQ card for differential voltage input and output and an IV preamplifier (SR570). Differential conductance measurements (dI/dV) are performed by a lock-in technique, which is realized by a LabVIEW code. Two computer controlled mechanical switches are used to disconnect the relatively noisy components of the circuit for conductance measurements and to connect the sample to the noise measurement circuit. The voltage noise across the sample is amplified by two sets of low noise amplifiers that are connected on both sides of the sample (NF Li-75a and Signal Recovery 5184). The amplified signals are collected by a dynamic signal analyzer (SR785), which calculates and averages the cross-spectrum of the two signals in order to cancel-out uncorrelated noise originating from the amplifiers. The amplifiers are powered by batteries in order to prevent exposure to noisy power lines. A differential voltage output of the low noise 4461 DAQ card is connected to the sample through two 500kΩ resistors to supply clean DC bias current to the junction during shot noise measurements. The system resides in a Faraday cage in order to reduce external electromagnetic noise. This version of the experimental setup is based on a previously introduced setup[1,2].

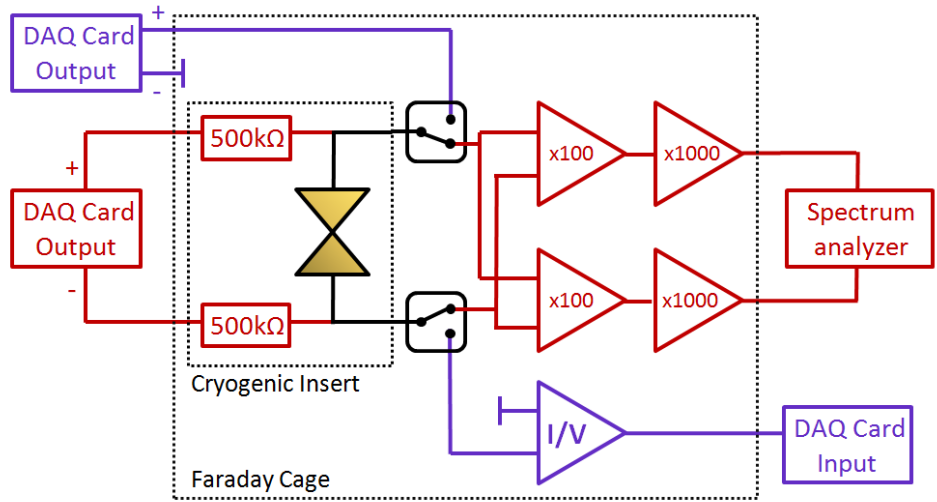

**Figure S1** – Electronic setup for conductance and shot noise measurements. Blue/red colored parts represent the conductance/noise measurement circuits, respectively. The dashed lined rectangles enclose the components which are located inside the cryogenic insert (inner rectangle) and Faraday cage (outer rectangle).

Shot noise measurements are performed by recording a set of noise power spectra as a function of bias voltage. An example set of spectra measured on an Au atomic contact is presented in Fig. S2(a). The thermal noise and shot noise originating from the sample are frequency independent in the measured range. In the presented spectra the signal is suppressed at high frequencies due to a low pass filter caused by the sample resistance and the parasitic capacitance of the cables as well as the input capacitance of the first set of amplifiers. The original signal is recovered by fitting a RC filter behavior to the zero-bias voltage spectrum and correcting the spectra according to the RC transfer function. The corrected spectra are cut to a selected frequency window (Fig. S2(b)). The excess noise is then calculated from the difference between the averaged noise power and the thermal noise power (at zero bias):

$$(1)\ S_{Iexcess} = \frac{S_V(V) - S_V(V=0)}{R^2}$$

where $S_V$ is the measured voltage noise power, $V$ is the bias voltage across the sample and $R$ is the resistance of the sample. Fig. 2(c) shows the bias voltage dependence of the excess noise power. The error is calculated from the standard deviation in $S_I$ values in the frequency window. The current noise of a ballistic conductor is given by[3]:

$$(2)\ S_I(V) = 4k_B T G_0 \sum_{n=1}^{N} \tau_n^2 + 2eV \coth\left(\frac{eV}{2k_B T}\right) GF$$

where $\tau_n$ are the transmissions coefficients of the N conduction channels, $F = \sum_{n=1}^{N} \tau_n(1-\tau_n)/\sum_{n=1}^{N} \tau_n$ is the Fano factor, and $k_B, G_0, T, G$ are the Boltzmann constant, the conductance quantum, the temperature and the conductance of the sample. At the low voltage limit ($eV \ll kT$), the expression is reduced to the Johnson-Nyquist expression for thermal noise $S_I = 4k_B TG$, while at high voltage ($eV \gg kT$), it has a linear dependence on bias voltage, $S_I = 2eVGF$. An example fit of this expression to the measured excess noise, assuming a single channel ($\tau_1 = 0.88$), is shown in Fig. S2(c) (red line).

In order to extract the Fano factor from the excess noise, we follow a procedure presented by Kumar et al. [2]. To simplify the fitting, two parameters are introduced:

$$(3)\ Y(V) = \frac{S_I(V) - S_I(0)}{S_I(0)}$$

$$(4)\ X(V) = \frac{eV}{2k_B T} \coth\left(\frac{eV}{2k_B T}\right)$$

Using these parameters, expression (2) is reduced to a linear relationship $Y(V) = [X(V) - 1]F$. Thus, the Fano factor is obtained from the slope of the linear fit to this expression (Fig. S2(d)).

A full shot noise measurement is performed in three stages. First, a dI/dV spectrum of a stabilized atomic configuration is measured as function of bias voltage. Next, the circuit is switched to the noise measurement configuration, and a series of noise spectra is measured as function of voltage applied on the sample. Finally, the circuit is switched back to conductance mode, and a second dI/dV spectrum is measured. The error in the conductance is calculated from the difference between the maximal and minimal conductance values recorded in both dI/dV spectra within the window of the shot noise measurement (0-5mV). Measurements with a conductance error larger than $0.02G_0$ are discarded.

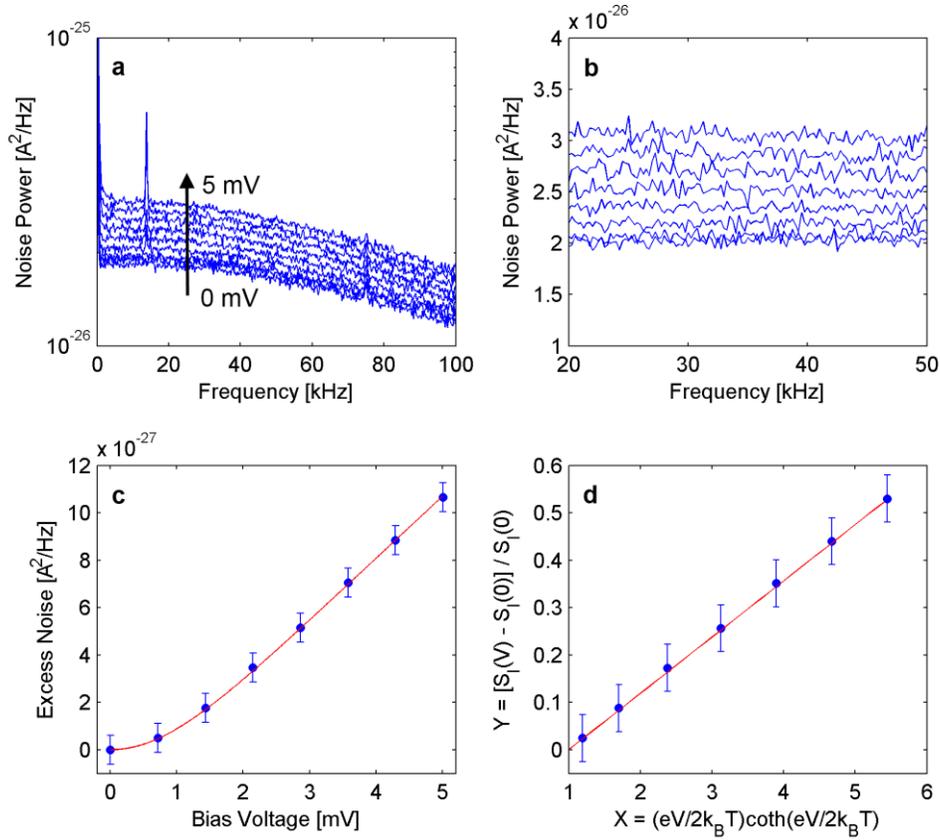

**Figure S2** – (a) Averaged spectra measured on a single-atom Au contact ($G=0.88G_0$) at 8 bias voltages values ranging from 0 to 5mV. Each presented spectrum consists of an average of 5000 successively measured spectra. (b) Corrected spectra within a selected frequency window of 20-50kHz. (c) Calculated excess noise as a function of bias voltage (blue circles), fitted by the theoretical expected noise for one channel ($\tau_1 = 0.88$). (d) The actual fit to the data is done using the excess noise, plotted for reduced parameters, X and Y. The best fitted line gives a Fano factor of F = 0.12±0.01.

## S2. Sensitivity of the computed transmission coefficients to Δτ

The presented numerical procedure enumerates the transmission coefficients using a fixed precision Δτ, which is limited by the computational effort needed for the enumeration. In order to verify that all possible solutions for $\tau_n$ are captured within the computed range $\tau_n^{min} - \tau_n^{max}$, we performed a sensitivity analysis of the procedure to Δτ. In this analysis, the computed $\tau_n^{min}$ and $\tau_n^{max}$ are examined as Δτ is reduced to increase the accuracy. Clearly, the reduction of Δτ results in more possible transmission sets $\{\tau_{i,n}\}_{n=1}^{N}$ that match the experimental values, and therefore this analysis can be used to check whether the additional solutions for $\tau_n$ do not exceed the initially obtained limits $\tau_n^{min}$ and $\tau_n^{max}$. Figure S3 shows the dependence of the computed range for $\tau_1$ as function of Δτ, for an example combination of G=1.00±0.01$G_0$ and F=0.14±0.02. The range is defined by $\tau_1^{min}$ and $\tau_1^{max}$ (marked as blue error bars) and it either decreases or keeps the same value for each reduction in Δτ. Therefore, for the given example, the new possible values for $\tau_1$ reached by lowering Δτ are already captured within the initial limits $\tau_n^{min}$ and $\tau_n^{max}$, which were computed for a higher Δτ.

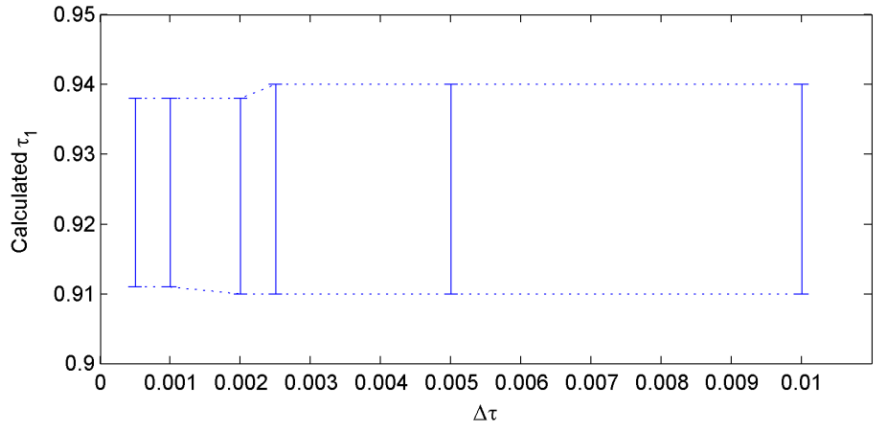

**Figure S3** – Computed ranges for $\boldsymbol{\tau_1}$ as function of Δτ, computed for G=1±0.01$G_0$ and F=0.14±0.02. Blue error bars represent the range of solutions limited by $\boldsymbol{\tau_1^{min}}$ and $\boldsymbol{\tau_1^{max}}$.

The above analysis was repeated for a large set of F and G values, which were uniformly spread within the full range of possible combinations for N=6: 0 ≤ F ≤ 1 and 0 ≤ G ≤ N, using intervals of 0.02 for F and 0.05 for G. For each combination, $\tau_n^{min}$ and $\tau_n^{max}$ were computed for three possible values of Δτ: 0.01, 0.005 and 0.0025 and for n between 1 to N. The analysis shows that for all considered cases, decreasing Δτ does not result in solutions that exceed the limits $\tau_n^{min}$ and $\tau_n^{max}$ computed for higher Δτ. This analysis indicates that $\tau_n^{min}$ and $\tau_n^{max}$ capture the whole range of solutions for $\tau_n$. A lower Δτ can improve the accuracy but will not add solutions outside the initial range of $\tau_n^{min}$ and $\tau_n^{max}$.

## S3. Sensitivity of the computed transmission coefficients to F and G

In this section, we apply our numerical approach to study the dependence of the transmission coefficients and the calculation uncertainty on the conductance (G) and Fano factor (F). The minimum and maximum of the possible coefficients, $\tau_n^{min}$ and $\tau_n^{max}$, are computed for the full range of F and G combinations, assuming a maximum of N=6 channels. The results for the first four coefficients $\tau_1 - \tau_4$ are plotted in Fig. S4. For each channel, we plot $\tau_n^{min}$, $\tau_n^{max}$ and the uncertainty, defined as $\tau_n^{max} - \tau_n^{min}$, as a function of F and G. The plots show that the possible F and G values for which $\{\tau_n\}_1^N$ have a solution are limited by a minimum and maximum Fano factor for each conductance value. The maximum Fano factor for N channels is given by $max(F) = 1 - G/N$, which is the case when all channels carry an equal transmission. The minimum Fano factor is reached when up to one channel is partly transmitting while the other channels are fully open ($\tau = 1$) or closed ($\tau = 0$), and is given by $min(F) = 1 - [\lfloor G \rfloor + (G - \lfloor G \rfloor)^2]/G$. An analytical proof for these limits can be found in[4].

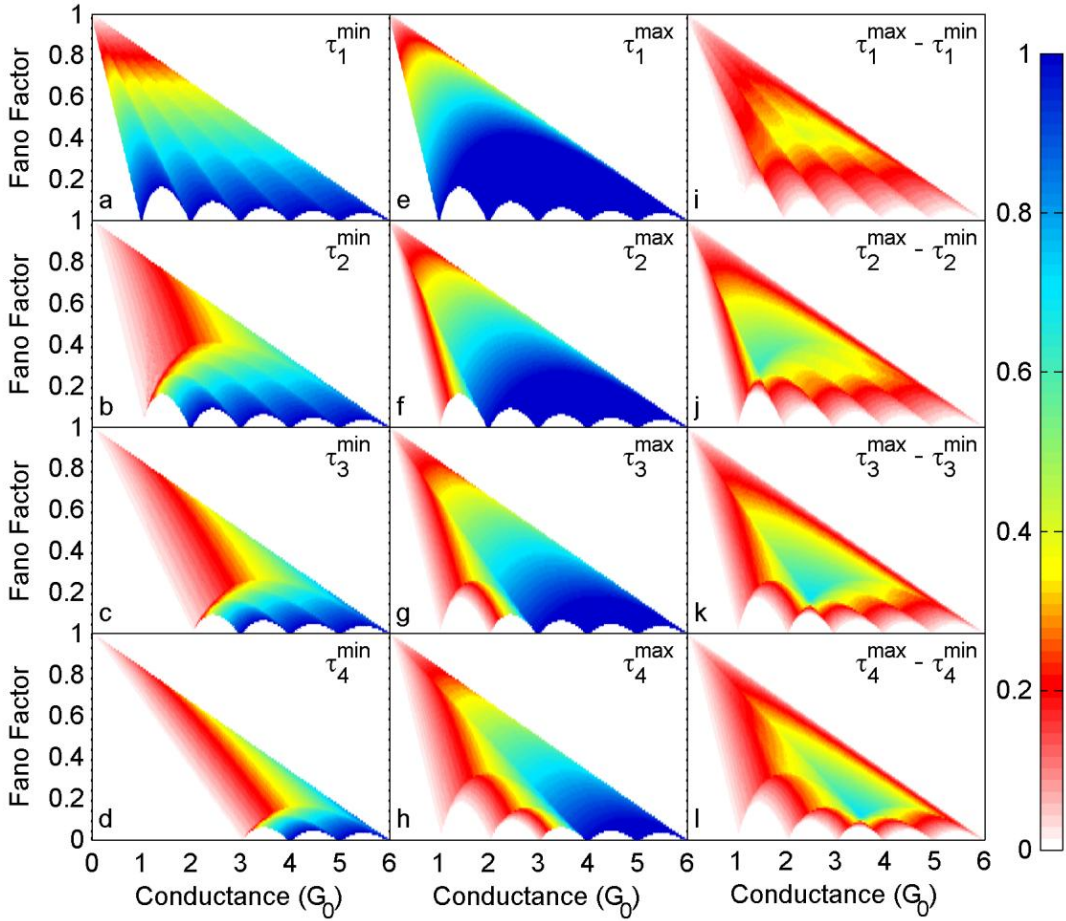

**Figure S4** – Computed $\tau_n^{min}$ (a-d) and $\tau_n^{max}$ (e-h) for the transmission coefficients as function of F and G and the uncertainty $\tau_n^{max} - \tau_n^{min}$ (i-l). The parameters used for the analysis are N=6, Δτ=0.01, ΔG=0.005 and ΔF=0.005. F and G were varied with intervals of 0.01 and 0.01$G_0$, respectively.

The plots of the uncertainty values indicate for which F and G combinations the transmission coefficients can be accurately determined. The plots show that, in general, the uncertainty is reduced near the limit of the minimum Fano factor. It can also be seen that the uncertainty for the highest transmitting channel $\tau_1$ is considerably lower than for the other channels. This implies that for most F and G combinations $\tau_1$ can be determined more accurately than the other coefficients. Another interesting observation is that the uncertainty for $\tau_1 - \tau_4$ increases significantly near $\min(F)$ at half integer conductance values starting from $1.5G_0$ (G = $1.5G_0$, $2.5G_0$, etc.), limiting the accuracy in determining $\tau_n$ for these F and G combinations. To demonstrate this effect, we calculate the transmission channels for a series of F and G combinations, where the conductance values are equally distributed in the range of $0.5 - 4G_0$ with an interval of $0.005G_0$. The simulated Fano factor values follow the minimum F curve according to $F' = \min(F) + 0.1$ (Fig. S5(a)). This example was chosen since it is related to the results for Ag contacts presented in the main article (Fig. 2(a)), in which more than 95% of the measured Fano factor values are found between $\min(F)$ and $F'$. Panels b-g in Fig. S5 show the dependence of the computed transmission coefficients on the conductance. As expected, the uncertainties of all six coefficients show to increase significantly at half integer conductance values. The simulated results demonstrate that the increased uncertainties obtained in the statistical analysis of Ag and Au contacts originate from the limited accuracy of the numerical method at these F and G combinations. For instance, in the case of $\tau_1$, once the value $\tau_1^{max}$ saturates, it remains equal to 1, while $\tau_1^{min}$ varies according to the uncertainty. Nevertheless, at conductance values of 2, 3 and $4G_0$, the low uncertainty enables to determine that the transmission of this channel is at least $\tau_1^{min} > 0.9$, suggesting that $\tau_1$ remains saturated to a large extent also at half integer values.

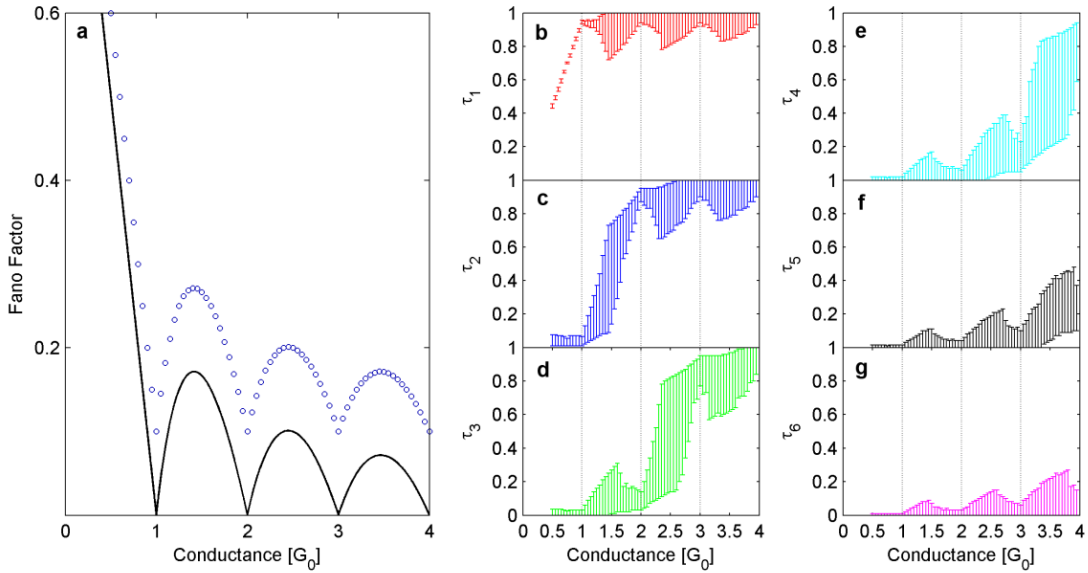

**Figure S5** – Computed transmission coefficients for a set of simulated F and G combinations presented in the text. (a) Simulated F and G values. (b-g) transmission ranges for τ₁-τ₆, computed using N=6, Δτ=0.01, ΔG=0.005 and ΔF=0.005.